# Physical simulation of resonant wave run-up on a beach

Alexander Ezersky[1]* and Nizar Abcha[1]

[1]UMR CNRS 6143 Morphodynamique Continentale et Côtière (M2C), Université Caen Basse Normandie, 24 rue des Tilleuls, 14000 Caen, France,

Efim Pelinovsky[2]

[2] Institute of Applied Physics, Russian Academy of Sciences, 46, Ul'janov St., Nizhny Novgorod 603950, Russia

Nonlinear wave run-up on the beach caused by harmonic wave maker located at some distance from the shore line is studied experimentally. It is revealed that under certain wave excitation frequencies a significant increase in run-up amplification is observed. It is found that this amplification is due to the excitation of resonant mode in the region between the shoreline and wave maker. Frequency and magnitude of the maximum amplification are in good correlation with the numerical calculation results represented in the paper (T.S. Stefanakis et al. PRL (2011)). These effects are very important for understanding the nature of rougue waves in the coastle zone.

It was revealed recently that the number of abnormally large and suddenly appearing waves (rogue waves) observed in the coastle zone is sufficiently large [1,2] The theory of the coastal rougue waves based on the nonlinear theory of shallow water [3-5]. To characterize the impact of waves on coastal infrastructure, the systematical study of run-up process is undertaken and a lot of papers summarizing the progress in the analytical solutions of the nonlinear shallow-water theory have been published by now (see for instance [7-15]). Recently, [6] on the basis of numerical simulations of the nonlinear shallow-water equations, it was pointed out the existence of resonance effects in the process of the long wave run-up. It should be noted that such resonance effect was predicted in [16] in the framework of linear theory.

The main result [6] is that at a certain frequency of the waves there exists a significant increase in the run-up amplitude. According to calculations, the maximal run-up height can be 50 times greater than the free surface oscillation amplitude used as the boundary conditions in the numerical calculations. It should be emphasized that the estimations presented in different papers cited above provide us an order of maximal run-up height much less than in [6]. It was established in [6] that the wave period for which maximal run-up amplification is observed depends on the slope of the bottom and the depth of water in the place where the waves are excited. This period is much larger than the "natural period" – time needed for perturbations to run from the point of excitation to the shoreline and return back. Results obtained in [6] pose a lot of questions. That is why in order to investigate the physical mechanisms of run-up amplification, , needed to explain the coastal rougue waves, we carried out a physical simulation of this process in the hydrodynamic channel with an inclined bottom. The simulation is carried out in such a way that its results can be used for comparison with numerical calculations presented in [6].

Before the presentation of experimental results, we give some theoretical estimates. The long wave run-up on a plane beach is described by the nonlinear shallow-water equations with constant coefficients:

$$\frac{\partial u}{\partial t} + u\frac{\partial u}{\partial x} + g\frac{\partial \eta}{\partial x} = 0,$$
$$\frac{\partial \eta}{\partial t} + \frac{\partial}{\partial x}(u(h+\eta)) = 0 \qquad (1)$$

$u$ is the depth-averaged velocity, $h=h(x)$ is the water depth, $\eta=\eta(x,t)$ is the free surface displacement, g is the acceleration of gravity. For linear growth water depth $h = x\theta$ ($\theta$ is the tangent of bottom slope angle), solutions of these equations may be found using the hodograph transformation (details can be found in papers [7-15]). For instance, if the incident wave far from the shore is a monochromatic wave of frequency $\omega$, the solution is presented in the following form:

$$\eta = \frac{1}{2g}\left(\frac{\partial \phi}{\partial \lambda} - u^2\right), u = \frac{1}{\sigma}\frac{\partial \phi}{\partial \sigma},$$
$$x = \frac{1}{2g\theta}\left(\frac{\partial \phi}{\partial \lambda} - u^2 - \frac{\sigma^2}{2}\right)$$
$$\phi(\sigma,\lambda) = \frac{2g\theta^2}{\omega}RJ_0(\omega\lambda/\theta g)\sin(\omega\lambda/\theta g) \qquad (2)$$

$\phi(\sigma,\lambda)$ is auxiliary function, $x$ is horizontal coordinate, and $J_0$ is the Bessel function of zeroth order. Generally speaking, wave field in the vicinity of the shoreline ($x = 0$) is not monochromatic, but if the nonlinear effects are ignored, the solution (2) is written in the flowing form :

$$\eta(x,t) = RJ_0\left(\sqrt{4\omega^2 x/g\theta}\right)\cos(\omega t) \qquad (3)$$

The constant $R$ plays role of the amplitude of the water oscillation on "unmoved" shore ($x = 0$). As it can be rigorously shown from (2), $R$ is the maximal run-up amplitude of the wave on the coast (moving shoreline). Far from the shoreline the wave is always linear because its amplitude is much less than water depth and it is a standing wave (3). Using asymptotic of the Bessel functions for large values of arguments we can select the incident wave with amplitude $A$ at the distance $L$ from the shoreline and obtain the amplification ratio in the following form [8,10,14]:

$$R/A = 2\sqrt{\pi}\left(h\omega^2/g\theta^2\right)^{1/4} = 2\pi\sqrt{2L/\lambda_0} \qquad (4)$$

where $\lambda_0$ is the wavelength on the isobath h located on distance $L$ from the shore. It is important to point that for large values of $L/\lambda_0$ the expression (4) presents the nonlinear amplification ratio and it can be obtained directly from (2). In the case of small values of $L/\lambda_0$ correct selection of incident and reflected waves is possible if the beach of length $L$ is matched with flat bottom. In this case the amplification ratio is given by [8,10,14, 17]:

$$R/A = 2/\sqrt{J_0^2(4\pi L/\lambda_0) + J_1^2(4\pi L/\lambda_0)} \qquad (5)$$

Comparison of (4) and (5) is given in Fig.1. As it can be seen the resonance effects are very small because the "resonator" ($0 < x < L$) is open from one boundary. For tsunami application, the characteristics of the incident wave very often are unknown. Meanwhile, a lot of buoys are now installed along the coasts (tide-gauge stations and DART buoys), and tsunami record on such buoys can be considered as input for solving of run-up problem. In the simplified geometry of plane beach, the buoy measures the standing wave (2) or (3), and in the case of small wave amplitude, free surface oscillations are described by:

$$\eta(L,t) = a\cos(\omega t), \quad a = RJ_0\left(\sqrt{4\omega^2 L/g\theta}\right) \qquad (6)$$

Amplification factor computed for the tsunami wave propagated from the buoy to the coast

$$C = R/a = J_0^{-1}\left(\sqrt{4\omega^2 L/g\theta}\right) \qquad (7)$$

has evident resonant properties and can be very large. Such big values of local amplification factors have been discovered in recent paper [6] where the result was obtained by numerical computation. This effect can be very important for prediction of tsunami behavior on the coast based on data of nearest buoy. Analytical and numerical results described above are examined below in physical experiment.

Experiments were performed in a long hydrodynamic flume 0.5 m width. The flume is equipped with a wave-maker controlled by computer. To simulate an inclined bottom a PVC plate with thickness of 1 cm is used. The plate is placed at different angles relatively to the horizontal bottom of the flume in the vicinity of the wave-maker (see Fig. 2). Two resistive probes are used to measure a displacement of water surface. Probes allow us to investigate the amplitude of free surface oscillations and phase of oscillation (one probe is used as a "clock") along the flume. Run–up height is determined by processing a movie which is shot by a high-speed camera mounted as shown in Fig. 2. Wave-maker allows us to excite harmonic wave of a given frequency and it works in two regimes: displacement-control and force-control. It is not possible to control free surface displacement, as it was done in the numerical experiment. That is why to study the ru-nup amplification, simultaneous measurements of the amplitude of free surface displacement near the wave maker and maximal run-up are carried out for different frequencies of excitation.

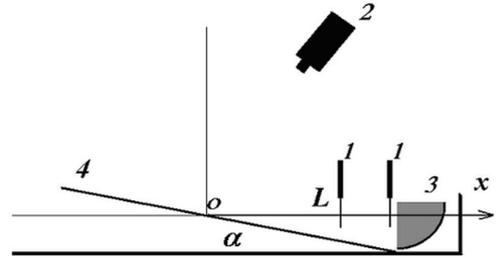

Fig. 2. Schema of experiment: resistive probes (1), high-speed video camera (2), wave maker (3), inclined bottom (4).

The frequency dependence of the amplitude of free surface displacement near the wave maker (*a*), maximal run-up (*R*) and coefficient of run-up amplification are shown in Fig. 3 for the slope of the bottom $\theta = tan(\alpha) = 0.263$. The amplitude of free surface displacement has peaks at frequencies $f_1 = 0.44$ Hz and $f_2 = 0.78$ Hz. They are the resonant frequencies of the system. The maximal run-up does have sharp picks, only a small increase of R in the vicinity of $f_1$ and $f_2$ is observed (Fig. 3b). The coefficient of run-up amplification (Fig. 3c) increases very sharply in the vicinity of $f_3 = 0.28$ Hz and $f_4 = 0.64$ Hz. It is evident that maximal amplification of run-up is observed for frequencies corresponding to the minimal amplitude (*a*). In the vicinity of the wavemaker the amplitude is sufficiently small and the signal is very noisy. That is why the coefficient of run-up amplification requires rather delicate measurements of free surface displacement: a band-pass filter was used to measure the amplitude of the harmonic corresponding to wave-maker forcing.

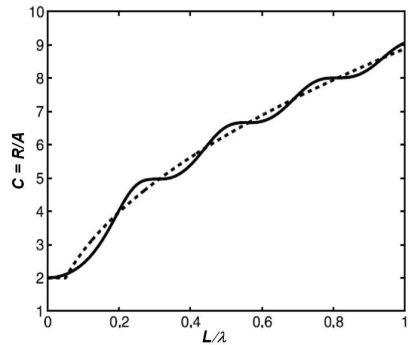

Fig. 1. Amplification run-up ratio: solid line obtained from equation (5), dashed line obtained equation (4).

It is important to note that the frequency of maximal amplification does depend on the method of wave excitation. Results presented in Fig. 3 were obtained for force-controlled regime of wave maker; the same results for coefficient of run-up amplification were obtained for displacement control regime.

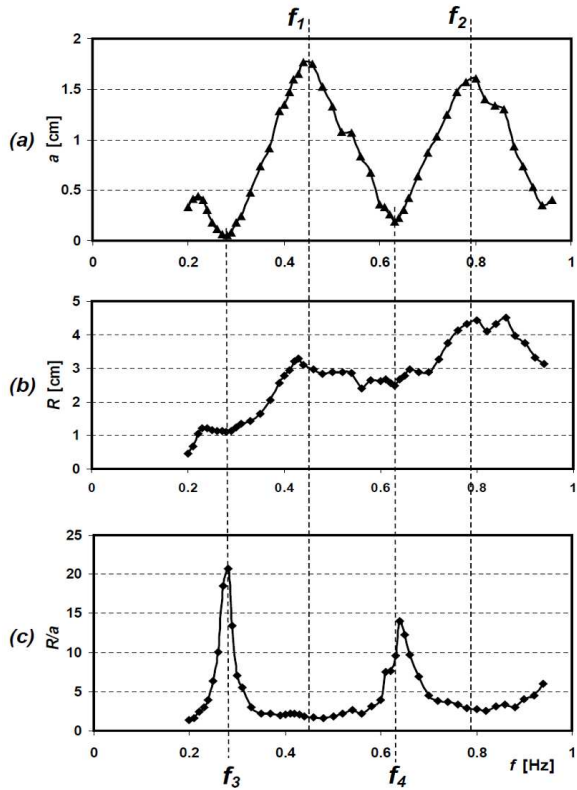

Fig. 3. Frequency dependence of amplitude of free surface displacement on frequency (resonance curve) (a), maximal run-up (b) and amplification of run-up (ratio of the maximal run-up and the amplitude of surface wave), (c) for slope $\tan\alpha = 0.263$.

Amplification coefficient $C$ was investigated for three bottom inclinations. Frequencies of maximal amplification depend on angle $\alpha$ and to compare results obtained for different angles $\alpha$, the non-dimensional frequency ($F$) was introduced: $F = f/f_0$, $f_0 = K^{-1}(g/h)^{1/2}\theta$, $K = 5.23$ ($h$ is for water depth near wave maker). Results are presented in Fig. 4. Non-dimensional frequencies of maximal run-up amplification $F_1 = 1$ for different angle $\alpha$ coincide very precisely. The coefficient of maximal amplification, corresponding to the frequency $F_1 = 1$ is approximately the same for different inclinations: $C \sim 20\text{-}25$. The second peak of run-up amplification coefficient is observed for frequency $F_2 = (2.2\text{-}2.3)F_1$. Non-dimensional frequency $F_2$ slightly depends on bottom slope; small peak is observed also for frequency $F_3 \sim 3.5\, F_1$.

It should be noted that in our experimental conditions, linear run–up (run-up without wave breaking) is observed for small frequencies of wave excitation $F < 2$, while for higher excitation frequencies $F > 2$ near $x = 0$ surface wave becomes strongly nonlinear and run-up occurs after the wave breaking. The wave breaking does not prevent precise determination of maximal run-up position. Excepting high frequencies $F > 3$, the border of the water on slope beach was one dimensional and maximal run-up did not depend on coordinate along direction perpendicular to axis $x$.

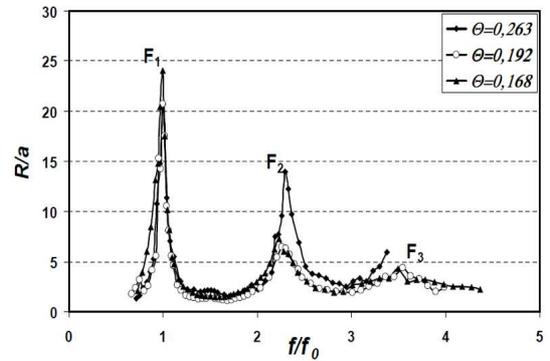

Fig. 4. Dependence of run-up amplification on normalized frequency for different bottom slopes.

To study frequency dependence of run-up amplification more precisely, the spatial structures of the free surface oscillations occurring at frequencies corresponding to the resonant frequencies of the system ($f_1$, $f_2$) and at frequencies of maximum run-up amplification ($f_3$, $f_4$) have been investigated. The results are shown in Fig. 5 for bottom slope $\theta = 0.168$. Amplitude and phase of free surface displacement are shown by rhombs and circles. Experimental data are compared with analytical solution (3) for free surface displacement $\eta$. Theoretical curves obtained from eq.(3) are shown in Fig. 4 by thick lines. The amplitude $a$ and the phase $\varphi$ of free surface harmonic oscillations are shown in Fig.4. Because Bessel function changes sing, amplitude is chosen as $a = |J_0|$, $\varphi = 0$ if $J_0 > 0$ and $\varphi = \pi$, if $J_0 < 0$. One can find in Fig. 5 that in the experiment the amplitude does not go to zero and phase changes smoothly for all frequencies. Note that frequencies of maximal run-up amplification ($f_3 = 0.205$ Hz, $f_4 = 0.46$ Hz) correspond to spatial modes having minimal amplitudes near the wave maker; resonance frequencies ($f_1$, $f_2$) have maximum amplitudes of free surface displacement near the wave maker. It should be noted that according to solution (3), frequencies of maximal run-up amplification correspond to the spatial modes with boundary condition $\eta(L,t) = 0$, and resonant frequencies correspond to mode with boundary conditions $\partial \eta(L,t)/\partial x = 0$. In other words, if one uses linear solution (1), the coefficient of run-up amplification in this approximation would be infinite: $a = 0$ at $x = L$. In the experiment, the amplitude is small, but finite. Comparison of curves presented in Fig. 5a,b,c,d shows that difference between theoretical solution and experimental data increases with frequency of excitation. For example, these differences are much more significant for $f_2$ than for $f_3$.

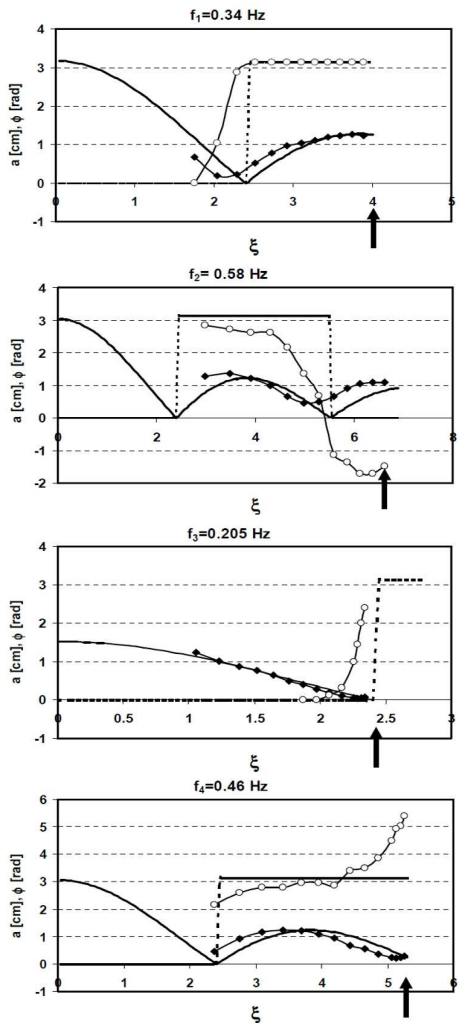

Fig. 5. Comparison of the experimental values of amplitude (rhombs) and phase (circles) with theoretical values of amplitude (thick solid lines) and phase (thick dash lines) obtained from the formula (3) $\xi = \sqrt{4\omega^2 x/g\theta}$, $\theta = 0.168$, arrows show the positions of the wave-maker edge.

Let us compare the experimental results with numerical simulations [6]. In the experiment, unlike the numerical calculations, it is not possible to introduce the waves with fixed free surface displacement at definite coordinate. In our opinion this point is not principal. Instead it, the simultaneous measurements of the free surface displacement and maximal run-up have been performed. In our experiment, the frequencies of maximal run-up amplification are very close to those that were obtained in the numerical calculation. We estimated the frequencies of the first peak as $f_3 = f_0 = K^{-1}(g/h)^{1/2}\theta$, $K = 5.23$; in [6] coefficient is estimated as $K \sim 5.1$. Second peak $f_4$ in the experimental frequency dependence of run-up is more visible than in numerical simulation [6]. Authors [1] did not give any estimations of the second frequency peak, but if one uses their data it is possible to conclude that frequency of the second peak is 2.5-2.7 times more than frequency of the first one. In our experiments the frequency of the second peak exceeds the frequency of first one in 2.2 – 2.3 times. Experimental values of frequencies $f_{3-4}$ practically coincide with frequencies of modes having nodes near the wave maker; numerical values [6] exceed this frequency by 2.5% for all bottom inclinations. The reason of such differences is not clear yet. Nonlinearity, wave dispersion, viscous dissipation influence the frequency of these peaks, but simple estimations for linear waves in shallow water with zero viscosity provide values which are close to experimental data. Authors [6] do not mention any dissipation of energy and non–linear parameter, which they use in numerical simulations. As for the coefficient of run-up amplification, the maximal value that was observed in the experiment is $C = 20\text{-}25$, whereas in [6] this value reaches $C = 50\text{-}60$. The difference is apparently due to viscous dissipation, which is essential in our experiments.

On the basis of experiments, we can conclude that the value of amplification coefficient and frequencies, at which run-up amplification maxima are observed, correlate with results of numerical simulations. The most important conclusion is that the existence of an abnormally large increase of the coefficient $C$ is due to resonator modes: this coefficient becomes very large because for its determination the amplitude at the mode node is taken as the amplitude of free surface displacement. This effect is very important for the explaination of rougue waves in the shore and for the prediction of tsunami run-up using the tide-gauge data. It is not sufficient to know the amplitude of free surface displacement in the nearshore zone. Each time it is necessary to know if this value corresponds to the amplitude $A$ of a propagating wave (eq. (4), (5)) or to the amplitude $a$ of a standing wave at a fixed point (eq.(6), (7)).

EP thanks a RFBR grant No. 11-05-00216.